# A FRAMEWORK FOR EVENT CO-OCCURRENCE DETECTION IN EVENT STREAMS

Laleh Jalali and Ramesh Jain
University of California, Irvine
*lalehj@uci.edu , jain@ics.uci.edu*

## ABSTRACT

This paper shows that characterizing co-occurrence between events is an important but non-trivial and neglected aspect of discovering potential causal relationships in multimedia event streams. First an introduction to the notion of event co-occurrence and its relation to co-occurrence pattern detection is given. Then a finite state automaton extended with a time model and event parameterization is introduced to convert high level co-occurrence pattern definition to its corresponding pattern matching automaton. Finally a processing algorithm is applied to count the occurrence frequency of a collection of patterns with only one pass through input event streams. The method proposed in this paper can be used for detecting co-occurrences between both events of one event stream (Auto co-occurrence), and events from multiple event streams (Cross co-occurrence). Some fundamental results concerning the characterization of event co-occurrence are presented in form of a visual co-occurrence matrix. Reusable causality rules can be extracted easily from co-occurrence matrix and fed into various analysis tools, such as recommendation systems and complex event processing systems for further analysis.

## 1. INTRODUCTION

Humans think in terms of events and entities. Events provide a natural abstraction of happenings in the real world. The concept of event is everywhere, from Lifelogs, multimedia experience sharing and video surveillance to healthcare. Multimedia research community was first focused on object based and entity based systems but now we are building applications that consider events as important as objects. Events need to be modeled effectively in order to be useful in over wide range of domains. Many event models have been introduced that focus on a variety of aspects of an event such as time, space, objects and persons involved [1,2]. However, causal, and correlative relationships between events have not been investigated in depth. In this paper we propose an event analytics framework that explores casual aspects between events by formulating co-occurrence relationship between events from multiple event streams.

Causality refers to the relationship between events where one set of events (the effects) is a consequence of another set of events (the causes). Causal inference is the process by which one can use data to make claims about causal relationships. Since inferring causal relationships is one of the central tasks of science, it is a topic that has been heavily debated in philosophy, statistics, and scientific disciplines. It's true that co-occurrence between events doesn't mean causality but once co-occurrence is analyzed and revealed through time, only then causality can be studied. Statistical correlation is not sufficient to demonstrate the presence of a temporal co-occurrence and causal relationship between events. So in our discussion, we use event co-occurrence term as finding *significant co-occurrence* relationship between event types over the time dimension. A significant co-occurrence between events means that one or more events co-occur within a specific time interval. So temporal constraints between events are the essence of co-occurrence definition. The principles described in this paper provide for identification of co-occurrences in *interval events*. Additionally, the principles provide for discovery of co-occurrences with arbitrary size, meaning that co-occurrences involve multiple (greater than or equal to two) event types from event streams. Figure 1 displays three event streams as the running example throughout the paper with $E_i \in \mathbb{S}$, i=1...6, $E'_j \in \mathbb{S}'$, j=1…4, and $E''_k \in \mathbb{S}''$, k=1…5. In this figure dash arrows represents co-occurrence between events. For example, suppose very often when $E_2$ occurs, then $E_1$ occurs *within 20 minutes* of the occurrence of $E_2$. It can be determined that $E_1$ and $E_2$ frequently co-occur given 20 minutes time constraint and this co-occurrence could be significant. Co-occurrence can be shown as a rule with the *cause-event* on the left side and the *effect-event* on the right.

In general, many applications consider events to happen instantaneously at one time point in time and apply time series analysis to detect relations between events [13,5]. As a result temporal reasoning will be limited to three binary relations for time points (before, equal, after). However, Events are perduring entities that unfold over time so they should be considered as a time interval. When dealing with time intervals the formulation of co-occurrence patterns is more complicated. Pattern detection in interval data has relied mostly on Allen's interval relations [9,11]. Yet, some researchers identified problems using Allen's relations such as ambiguousness in the pattern representation because the same pattern can describe different situations in the data, and lack of robustness to noise because small shifts of time points lead to different patterns [10]. In this work we propose using semi-intervals [12] to define more flexible patterns. For example in figure 1, in the second and third occurrence of $E_1$, $E_1$.start is unknown while $E_1$.end is known. Hence, the pattern $E_1 \xrightarrow{15} E'_3$ is ambiguous since we haven't stated whether occurrence of $E'_3$ should be within 15 time units of $E_1$.start or $E_1$.end. The same thing is true for $E_2 \xrightarrow{20} E_1$. One way to solve this ambiguity is allowing a mixture of intervals and semi-intervals in pattern definitions. Using semi-intervals enables one to capture more pattern cases in the data as opposed to complete intervals. We explain the formal definition of such patterns in the next section.

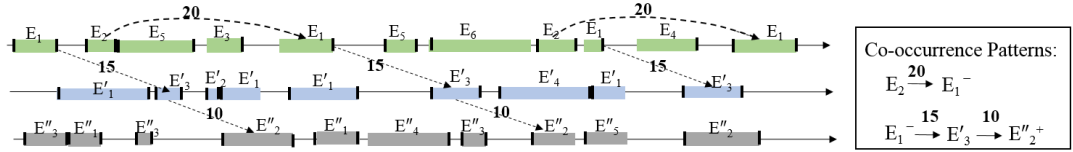

**Fig. 1.** Three event streams and sample events with co-occurrence relations between them. Bold lines indicate the interval boundaries for each event. In some events one of the boundaries might be missing so the bold line is not depicted for that event.

Complex Event Processing (CEP) research is primarily focused on pattern matching techniques over real-time event streams (e.g., Cayuga [4], SASE [13], ZStream [3]). Our approach is different than CEP systems since we are working on archived event streams to support longer-term event analytics. Temporal reasoning and relational ordering of event data has been extensively researched. However, not as much research has been focused on processing using semi-intervals. Situated within the data mining domain, the symbolic temporal pattern mining approach focuses on discovering frequent patterns among symbolic time series data [6,7]. One such approach is T-patterns developed by Magnusson [8] in which a sequence of events will occur within certain time windows of each other and time windows are set through various statistical methods. Our work is different since we are not looking for frequent patterns or predicting patterns, but the interest of our proposed framework is understanding how do events relate structurally and temporally, and how characterizing the co-occurrence relationships between events helps in understanding *potential* causal relationships between them.

In this paper we present an event analytics framework for co-occurrence detection from multiple event streams where events are represented as semi-intervals. Our approach is different from most event correlation techniques that consider events to be instantaneous point in time. Our approach is novel because patterns can be a mixture of intervals and semi-intervals. So we can define more patterns in the event streams as opposed to interval-based or instantaneous point patterns. We present a computational algorithm that counts the occurrences of a collection of patterns from multiple event streams with just one pass through the data. Also we demonstrate the result of the analysis as a co-occurrence matrix, a powerful technique for co-occurrence visualization between two event streams through time. Co-occurrence rules extracted from such processing can be used further to study causal relationships between events.

## 2. EVENT CO-OCCURRENCE FRAMEWORK

Our event analytics framework for co-occurrences detection is based on creating a formalism for co-occurrence pattern structure with temporal constraints and ordered relationships between events, translating the pattern to a corresponding finite state automaton, and a processing algorithm that counts occurrences of patterns in input event streams. Patterns can include complete intervals or only the starting and ending time points expressing a mixture of intervals and semi-intervals. Semi-intervals allow a flexible representation where partial or incomplete knowledge can be handled since operation is on parts of an interval and not the whole. In [10] the use of interval boundary representation was proposed for mining Allen's relationships where the TPrefixSpan algorithm mines frequent patterns composed of complete intervals. We propose a new definition for semi-interval patterns using the same data representation as [10] but the class of patterns defined and the processing algorithms are completely different.

In this section we discuss how semi-intervals are used to describe events, event streams and patterns. Then we explain how a high-level definition of a pattern with implicit structural and temporal information can be translated into automata-base pattern specification using automaton building blocks.

### 2.1. Definitions

We will gradually establish our semantics to meet the above-mentioned goal, beginning with a number of fundamental concepts (such as event streams, pattern definitions, co-occurrence, etc.). The input to our framework is a set of event streams. Each event stream contains events of different types. The process of generating event streams from data streams is not in the scope of this paper and has been a popular topic in multimedia community for video and audio event detection. Event types from the same event stream conforming to a specific ontology that defines the vocabulary $\Sigma$ for labeling those event types.

*Definition 1* (Time Domain). A time domain T is a discrete, linearly ordered, countably infinite set of time instants t ∈ T. We assume that T is bounded in the past, but not necessarily in the future.

*Definition 2* (Time Interval). A time interval is a tuple where $[t_s, t_e] \in T^2$ and $t_s \leq t_e$. The finite set of all time intervals is noted $I = \{ [t_s, t_e] | t_s \leq t_e \}$

*Definition 3*. We define an order relations ≺ as follows:
$$T \prec T' \equiv (t_s < t'_s) \vee (t_s = t'_s \wedge t_e < t'_e)$$

*Definition 4*. A labeled time interval is a triple $[\partial^{+/-}, t_s, t_e]$ where $\partial \in \Sigma$ is a unique symbol, and interval boundaries are represented with + and − signs where $\partial^+$ and $\partial^-$ are corresponding to start and end of the interval respectively.

*Definition 5* (Event). An event $e \langle v, [\partial^{+/-}, t_s, t_e] \rangle$ consists of a tuple $v$ conforming to a schema €, with a start time value $t_s$ and end time value $t_e$. $\partial$ is label of the event. Events with complete interval are represented as $(\partial, t_s, t_e)$, while events with semi-interval are represented as $(\partial^+, t_s)$ when $e.t_s$ is available, and $(\partial^-, t_e)$ when $e.t_e$ is available.

*Definition 6* (Event Stream). A stream 𝕊 is a totally ordered, countably infinite sequence of events such that:
$$\forall e_i, e_j \in S, e_i \prec e_j, iff (e_i.t_s < e_j.t_s \vee (e_i.t_s = e_j.t_s \wedge e_i.t_e < e_j.t_e))$$
Event stream 𝕊 shown in figure 1 is encoded as follow:
($E_1$,1,5) ($E_2$,8,11) ($E_5$,11,18) ($E_3^-$,22) ($E_1^-$,30) ($E_5$,35,40) ($E_6^+$,42) ( $E_2$,53,57) ( $E_1^-$,60) ($E_4^-$,71) ($E_1$,73,76)

Next we define co-occurrence patterns that encode temporal and structural relationships between events. Time constraint information is critical in co-occurrence detection. This constraint is specified by giving an interval of the form $\Delta t = [\delta_1, \delta_2]$ and requires the difference between the times of every pair of successive events in any occurrence of a pattern to be in this interval. Formal definition of a pattern is as follow:

*Definition 7* (Co-occurrence Pattern). Specify a particular order in which the events of interest should occur. It, however, allows an arbitrary number of events to appear between the two events addressed by two consecutive parameters. This pattern is defined recursively as a recurring ordered set:
$$P = \left( \left( \left( E_1^{+/-}; E_2^{+/-} \right)_{\Delta t1}; E_3^{+/-} \right)_{\Delta t2} ...; E_m^{+/-} \right)_{\Delta tm-1}$$
Can also be written as:
$$P = E_1^{+/-} \xrightarrow{\Delta t1} E_2^{+/-} \xrightarrow{\Delta t2} E_3^{+/-} ... \xrightarrow{\Delta tm-1} E_m^{+/-}$$
Size of the pattern is equal to the number of participating events. The general term $(E_i^{+/-}; E_j^{+/-})_{\Delta t}$ reads: in the occurrence of the pattern, when $E_i$ begins/ends, then within $\Delta t$ time units later $E_j$ begins/ends. The pattern structure consists of an ordered set of intervals as well as an ordered set of event types. The time difference between successive events in any occurrence has to be in the prescribed interval. For example the pattern $\rho 1 = (E_2; E_1^-)_{20}$, reads as *when $E_2$ happens within 20 time unites $E_1$ ends*, has two occurrences in 𝕊: {($E_2$,8,11)($E_1^-$,30)} and {( $E_2$,53,57)($E_1^-$,60)}.

Pattern $\rho 1$ is defined between events from the same event stream. In general, we are interested in finding temporal patterns between events from two or more streams. Events from multiple streams might be overlapping so in co-occurrence detection we shall preserve temporal ordering between them. To do so we apply temporal sequence arrangement rules from [10] to *serialize* events from two or more streams and generate one serialized event stream containing all event types from those streams. The graphical representation and encoding process are demonstrated in figure 2. The higher-level pattern language specified above, is translated to a corresponding pattern matching automaton. The automaton is then used in the processing algorithm to count the frequency of pattern in the input event stream.

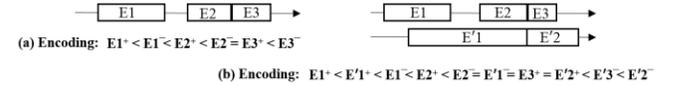

(a) Encoding: $E_1^+ < E_1^- < E_2^+ < E_2^- < E_3^+ < E_3^-$

(b) Encoding: $E_1^+ < E'_1^+ < E_1^- < E_2^+ < E_2^- < E'_1^- < E_3^+ < E'_2^+ < E_3^- < E'_2^-$

**Fig. 2.** (a) Example encoding of a sequence of events. $E^+_1$ and $E^-_1$ represent the start and end times of event $E_1$, respectively. Relational operators are used to indicate the ordered relations between start/end times. (b) Example of serializing events from two event streams.

**2.3. Structure of Automata**

Co-occurrence pattern detector in our framework employs a new type of automaton that comprises Finite State Automata (FSA) with support for a rich time model and event parameterization thus called $FSA^f$. This automaton contains a finite number of states and state transitions (edges), but each edge also maintains the timing information of the previous event detected. One automaton is created and initialized for each pattern. Formally an automaton $FSA^f$ = (OS, TS, E, θ, F), consists of a set of ordinary states OS, a set of time states TS, a set of directed edges E, a set of formulas θ labeling those edges, and a final state F. These automaton building blocks are shown in figure 3(a). In a given state, the automaton decides when to make a transition to another state by evaluating a formula associated with each edge. The automaton that encodes pattern ρ = $E_A^+ \xrightarrow{\Delta t1} E_B^- \xrightarrow{\Delta t2} E_C$ is demonstrated in figure 3(b). The basic ingredient of computational algorithm is $FSA^f$ that is used to recognize (or track) a pattern's occurrence in the event stream.

**States**: our automaton has two types of state: ordinary state and time state. In general, a pattern with size N has 2N states in $FSA^f_\rho$ (N ordinary states, N-1 time states and a final state). Ordinary state is represented by a pair (i,$OS_\rho[i+1]$) where i=0,…N-1, means that the $FSA^f$ has already seen the first i event types of this pattern and is waiting for $OS_\rho[i+1]$. If we now encounter an event of type $OS_\rho[i+1]$ in the event stream, it validates the formula on the $\theta_{s\_edge}$ edge. If the condition in the formula is satisfied it proceeds to the next state. A time state is represented by $TS_\rho[\Delta t_i]$ where $\Delta t_i = [\delta_1, \delta_2]_i$ is the time interval indicated in the pattern formulation. $\Delta t$ value is used to initialize time interval boundaries $\delta_1$ and $δ_2$ which will be used later in the automaton edge formula of the next state to progress or fail the match. The start state of the FSA$^f$ is (0, $OS_ρ[1]$). The final state or the $2N^{th}$ state is (N, ϕ) where ϕ is a null symbol. The final state is the accepting state because when the automaton reaches this state, a full occurrence of the pattern is tracked.

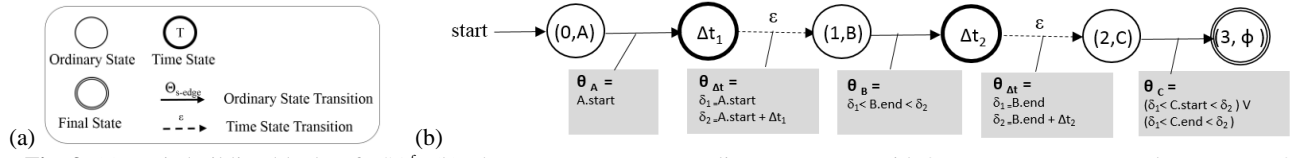

**Fig. 3.** (a) Basic building blocks of FSA$^f$. (b) The automaton corresponding to pattern ρ with 3 event components. It demonstrates 3 ordinary states, 2 time states, and formula associated with each state transition edge.

**Edges**: each state is associated with an outgoing transition edge representing the action that can be taken at the state. Each edge at a state *s* is precisely described by a **formula** that specifies the condition on taking it denoted by $θ_{s\_edge}$. Formula of each edge is compiled from the pattern specification. We use solid lines to denote the outgoing edge from ordinary state. This edge consumes an event from the input and validates its formula and tries proceeding. The dash line out of time state is a special ε-edge, it doesn't consume any input event but only initialize time interval boundaries $δ_1$ and $δ_2$. For example pattern ρ needs an start occurrence of event A ($E_A^+$) followed within $Δt_1$ time units with and end of event B ($E_B^-$) followed within $Δt_2$ time units with either start or end of $E_C$. $θ_A$ evaluates whether start of event A happens or not. $θ_{Δt1}$ initialize time constraint boundaries that will be used in $θ_B$ to evaluate whether end of event B happens within the designated boundaries. Same process happens for $θ_{Δt2}$ and $θ_C$ with a difference that according to pattern ρ's definition either start or end of event C should be within the designated boundaries. As we shall see later, all processing algorithms are obtained by different ways of managing a set of such automata.

### 2.4. Auto Co-occurrence and Cross Co-occurrence

*Definition 8* (Auto Co-occurrence). For a pair of events $E_i$, $E_j$ belongs to the same stream, $E_i \in \mathbb{S}$ and $E_j \in \mathbb{S}$, Auto co-occurrence with *temporal offset* Δt is the frequency count of the pattern $(E_i^{+/-}; E_j^{+/-})_{Δt}$ normalized by the frequency of $E_i$ in the stream.

$$AutoCo-occurrence(E_i, E_j, Δt) = \frac{count((E_i^{+/-}; E_j^{+/-})_{Δt})}{count(E_i)}$$

This definition means that from all the times that $E_i$ happens, how many times it's followed by $E_j$ within Δt time units.

*Definition 9* (Cross Co-occurrence). For a pair of events $E_i$, $E'_j$ belong to different event streams $E_i \in \mathbb{S}$, $E'_j \in \mathbb{S}'$, cross co-occurrence with *temporal offset* Δt is the frequency count of the pattern $(E_i^{+/-}; E'^{+/-}_j)_{Δt}$ normalized by the frequency of $E_i$ in the stream.

$$CrossCo-occurrence(E_i, E'_j, Δt) = \frac{count((E_i^{+/-}; E'^{+/-}_j)_{Δt})}{count(E_i)}$$

## 3. COMPUTATIONAL ALGORITHM

The purpose of this study is to detect co-occurrence between events from multiple event streams. To do so, we first need to model event co-occurrences as structural and temporal co-occurrence patterns and then apply processing algorithm to detect frequencies of these patterns from input event streams. In this section we explain how the processing algorithm count the occurrences of patterns using FSA$^f$.

The strategy for counting occurrences of a pattern is straight forward. For a pattern, say ρ, an automaton FSA$^f_ρ$ is initialized at the earliest event in the event stream that corresponds to the first event of ρ. The initialization process includes translating event types and temporal constraints to ordinary states and time states, assigning formula $θ_{s\_edge}$ to each outgoing edge, and allocating a buffer to store $δ_1$ and $δ_2$ values as well as the formula. Due to space limitations we omit a detailed description of initialization process. As we read data from event stream, by evaluating each edge's formula this automaton makes earliest possible transitions into each successive state. Once it reaches its final state, an occurrence of the pattern is recognized and its frequency is increased by one. A fresh automaton is initialized for this pattern when an event corresponding to its first event appears again in the event stream and the process of recognizing an occurrence is repeated. Algorithm 1 gives the pseudo code for the processing algorithm. The set of co-occurrence patterns of interest and a serialized event stream are needed as input. The output of the algorithm is the frequency count of each input pattern. To access and traverse the automata efficiently they are indexed using a *wait(.)* list where for each event type *E*, the automata that accept E are linked together to a list *waits(E)*. The list contains entries of the form (ρ, x) meaning that pattern ρ is waiting for its $x^{th}$ event. This idea of efficiently indexing automata through a *waits(·)* list was introduced in the windows-based frequency counting algorithm [14] with the difference that instead of simple Finite State Automata we introduce FSA$^f$ extended with a rich time model and support for event parameterization. For pattern ρ of size N, we have a list of ordinary states $OS_ρ[i]$, i=1..N and a list of time states $TS_ρ[i]$, i=1..N-1. The *waits(·)* lists are initialized by adding the pair (ρ, 1) to waits($OS_ρ[1]$), for each pattern ρ. The main loop in the algorithm looks at each event in the input stream and makes necessary changes to the automata in *waits(·)*. When processing the $i^{th}$ event in the serialized event stream, the automata in *waits($E_i$)* are considered. Every automaton waiting for $E_i$ is transited to its next time state if formula $\theta_{s\_edge}$ is satisfied. Then $\delta_1$ and $\delta_2$ values are initialized so they can be used in the next edge's formula and automaton proceeds to the next ordinary state. If the automaton has not yet reached its final state, it waits next for ($\rho$, x + 1). If automaton has reached its final state, then a new automaton for the pattern is initialized by adding ($\rho$, 1) to waits($\rho$[1]) and the frequency of the pattern is increased by one.

---

**Algorithm 1** Counting frequency of co-occurrence patterns
**Input:** Serialized event stream S = {($E_1^{+/-}$,$t_1$), ($E_2^{+/-}$,$t_2$),…($E_n^{+/-}$,$t_n$)}, Set P of sequential patterns of interest
**Output:** The set F of frequent count of each pattern from P
1: for all event types E in S
2:   Initialize waits(E) = φ
3: for all $\rho \in$ P do
4:     Initialize an automaton $FSA_\rho^f$ = = ($OS_\rho$, $TS_\rho$, $E_\rho$, $\theta_\rho$, $F_\rho$)
5:   Initialize $\rho$.freq = 0
6:   Add ($\rho$,1) to waits($OS_\rho$[1])
7: for i = 1 to n do / ∗ n is length of event stream ∗/
8:   for all ($\rho$ , x) ∈ waits($E_i$) do
9:       If ($\theta_\rho$ is valid)
10:         Proceed to $TS_\rho$[x]
11:           Set x′=x+1
12:           Initialize $\delta_1$ and $\delta_2$
13:         Proceed to $OS_\rho$[x′]
14:         Remove ($\rho$,x) from waits($E_i$)
15:         if x′ = (N+1) then
16:             Set x′=1
17:         Add ($\rho$,x′) to waits($OS_\rho$[x′])
18:         If x = N then
19:             Update $\rho$.freq = $\rho$.freq + 1
20: Output F = { $\rho$.freq | $\rho$ ∈ P }

---

## 4. EVALUATION RESULTS

This section presents the results obtained from some synthetic data that generated by embedding specific interval and semi-interval patterns in varying levels of noise. The main objective of the experiments is to empirically demonstrate the ability of our event analytics framework in detecting co-occurrence patterns with different sizes. By varying the control parameters of synthetic data generation, it's possible to generate qualitatively different kinds of datasets.

Each semi-interval pattern to be embedded in the synthetically generated data, consists of a specific ordered sequence of events and time constraints between them. Data generation process is as follow: There is a timer that specifies the current time instant. Each time an event is generated, this timer specifies event's start time. The duration of each event is picked from a normal distribution with a mean value $\mu$. Event's end time is the sum of its start time and duration. Number of event types defined by $|\partial|$ and n is the number of events in the stream. Minimum temporal granularity is set to 1 min. However, this value is application dependent and can be as small as millisecond. After generating an event, timer is incremented with a small random integer. Each time the next event is to be generated, two decisions should be made. 1) Whether the event is going to have both start and end timestamps, or one of them might be missing randomly. This is controlled by the parameter $\alpha$, which is the probability that the next event has its both interval boundaries. If $\alpha$ is 1 then event stream contains only complete intervals. 2) Whether the next event is to be generated randomly with uniform distribution over all event types or according to one of temporal patterns to be embedded. This is controlled by the parameter $\beta$, which is the probability that the next event is generated randomly. If $\beta$ is 1 then data contains only noise with no temporal patterns embedded. If it decides that the next event is to be from one of the temporal patterns to be embedded, then we have a choice of continuing with a pattern that is already embedded partially or starting a new pattern. If time constraints of the partial pattern cannot be satisfied any more, we start a new occurrence of the pattern. Five datasets with varying amount of noise are generated. We embed the following pattern of size six in all datasets: $\rho = E_A^+ \xrightarrow{15} E_B \xrightarrow{10} E_C^- \xrightarrow{20} E_G \xrightarrow{60} E_H \xrightarrow{90} E_D$

Data generation with $\beta$=0.2 means that with 20% probability the next event is generated randomly and with 80% probability either pattern $\rho$ is continued or a new occurrence of $\rho$ is started. Frequency count of pattern $\rho$ from size one ($E_A^+$), size two (($E_A^+$; $E_B$)$_{15}$), to size six (complete pattern) is plotted in figure 4. Our objective is to see whether this pattern can be detected based on its frequency counts given different amount of noise in the data.

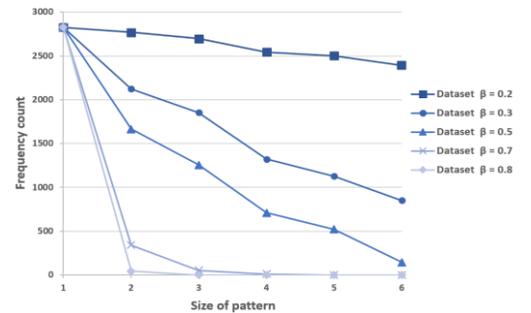

**Fig. 4.** Frequencies of pattern $\rho$ with sizes from 1 to 6 in five dataset with n=$10^6$, $|\partial|$ =22, $\alpha$=0.3, and varying amount of noise $\beta$. Dataset with $\beta$=0.2 has the least noise.

It's apparent that when the noise increases, frequencies of patterns (partial patterns and complete pattern) falls quickly. Looking at the curves corresponding to five datasets we see that decrease of pattern's frequency with increasing size is directly related to how much noise is injected in the dataset. Thus we can say that long patterns with high frequencies cannot come out unless there are strong co-occurrence connection between corresponding event components of the pattern in the underlining data generation model.

In image processing a co-occurrence matrix or co-occurrence distribution is a matrix that is defined over an image to be the distribution of co-occurring values at a given offset to measure the texture of the image. In our work, we use co-occurrence matrix in a different way. As shown in figure 5, *x* and *y* axes of the matrix are composed of event types from input streams. In case of measuring auto co-occurrence, both axes contains the same event types. In case of measuring cross co-occurrence, each axes contains event types from one of the streams. Note that $|\partial|=N$ will yield to an N by N co-occurrence matrix for a given $\Delta t$. Each cell of the matrix C is the co-occurrence value calculated from definition 8 (or 9) for a pair of events:

$$C_{\Delta t}(i,j) = Co-occurrence(E_i, E_j, \Delta t), \forall i,j=1...N$$

Co-occurrence matrix can only visualize patterns of size two. By changing temporal offset, multiple co-occurrence matrices can be computed. In this experiment we generate a dataset with $n=10^5$, $|\partial|=22$, $\beta=.8$, and three patterns are embedded $(E_C; E_F)_{15}$, $(E_I; E_M)_{30}$, and $(E_S; E_H)_{60}$. Three co-occurrence matrices are demonstrated in Figure 5. Such a visualization facilitates browsing co-occurrence characteristics in event streams, formulating hypothesis regarding those characteristics, and investigating potential causal relationships between multimedia events.

## 5. CONCLUSION

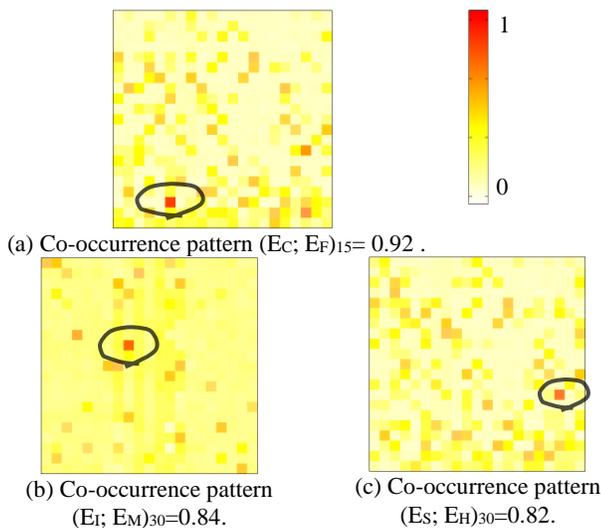

(a) Co-occurrence pattern $(E_C; E_F)_{15}= 0.92$.

(b) Co-occurrence pattern $(E_I; E_M)_{30}=0.84$.

(c) Co-occurrence pattern $(E_S; E_H)_{30}=0.82$.

**Fig. 5.** Co-occurrence matrices with different temporal offsets. (a) $\Delta t=15$ min, (b) $\Delta t=30$ min, (c) $\Delta t=60$ min.

Events are an important concept in multimedia. Tremendous research conducted in the past for recognizing events from data streams but we lack a rigorous framework to build causality models using event analytic techniques. In this paper, we show co-occurrence analysis is the first yet very powerful step towards understanding *potential* causal relations. The characterization and efficient representation of this relation is a non-trivial problem. So the formulation presented in this paper will be useful for reasoning in visual surveillance domain, video understanding, and emerging area of wearable devices. We are applying this framework to wellness applications, particularly related to asthma management to detect sensitivity of an individual to her activity-level, pollution, and pollen.

## 6. REFERENCES

[1] Westermann, Utz, and Ramesh Jain. "Toward a common event model for multimedia applications." *IEEE MultiMedia*, 14(1): 19-29, 2007.
[2] Scherp, Ansgar, et al. "F--a model of events based on the foundational ontology dolce+ DnS ultralight." *Proc. of the 5th International Conference on Knowledge Capture*. ACM, 2009.
[3] Mei, Yuan, and Samuel Madden. "Zstream: a cost-based query processor for adaptively detecting composite events." *Proc. of the 2009 ACM SIGMOD International Conference on Management of data*. ACM, 2009.
[4] Demers, Alan J., et al. "Cayuga: A General Purpose Event Monitoring System." *CIDR*. Vol. 7, 2007.
[5] Shumway, Robert H., David S. Stoffer, and David S. Stoffer. "Time series analysis and its applications." *Vol. 3. New York: Springer*, 2000.
[6] S. Laxman and P. Sastry. "A survey of temporal data mining." *Sadhana*, 31(2):173–198, 2006.
[7] F. Mörchen. "Unsupervised pattern mining from symbolic temporal data." *ACM SIGKDD Explorations Newsletter*, 9(1):41–55, 2007.
[8] M. Magnusson. "Discovering hidden time patterns in behavior: T-patterns and their detection." *Behavior Research Methods*, 32(1):93–110, 2000.
[9] Laxman, Srivatsan, and P. Shanti Sastry. "A survey of temporal data mining." *Sadhana* 31.2 (2006): 173-198.
[10] Wu, Shin-Yi, and Yen-Liang Chen. "Mining nonambiguous temporal patterns for interval-based events." *IEEE Transactions on Knowledge and Data Engineering,* 19(6): 742-758, 2007.
[11] Allen, James F. "Maintaining knowledge about temporal intervals." *Communications of the ACM*, 26(11):832-843, 1983.
[12] Freksa, Christian. "Temporal reasoning based on semi-intervals." *Artificial intelligence,* 54(1): 199-227, 1992.
[13] Gyllstrom, Daniel, et al. "SASE: Complex event processing over streams." *3rd Biennial Conference on Innovative Data Systems Research (CIDR)*, 2006.
[14] Mannila, Heikki, Hannu Toivonen, and A. Inkeri Verkamo. "Discovery of frequent episodes in event sequences." *Data Mining and Knowledge Discovery*, 1(3): 259-289, 1997.